\documentclass[final,5p,times,twocolumn]{elsarticle}

\usepackage[utf8]{inputenc}
\usepackage{amsmath,amssymb,amsfonts}
\usepackage{bm}
\usepackage{graphicx}
\usepackage{xcolor}
\usepackage{booktabs}
\usepackage[colorlinks=true,breaklinks=true,linkcolor=blue,citecolor=purple,urlcolor=teal]{hyperref}
\usepackage{url}

\renewcommand{\Im}{\mathop{\mathrm{Im}}}

\journal{Physics Letters B}

\biboptions{comma,sort&compress}

\begin{document}
\begin{frontmatter}

\title{Exact Treatment of Continuum Couplings in Nuclear Optical Potentials via Feshbach Theory}

\author[tongji]{Hao Liu}
\author[tongji]{Jin Lei\corref{cor1}}
\cortext[cor1]{Corresponding author}
\ead{jinl@tongji.edu.cn}
\author[tongji]{Zhongzhou Ren}
\address[tongji]{School of Physics Science and Engineering, Tongji University, Shanghai 200092, China}

\begin{abstract}
We present a full-coupling construction of the Feshbach effective interaction in a converged continuum-discretized coupled-channels (CDCC) calculation. The method retains the complete Green's function in the excluded continuum space, treating the continuum-continuum couplings to all orders within the discretized CDCC model space, and therefore yields an explicitly non-local dynamic polarization potential. Applied to $d+^{58}$Ni scattering, the projected two-body potential reproduces the parent CDCC elastic observables and gives a reasonable description of the available experimental data, providing a direct numerical check of the projection. The resulting non-local potential exposes the spatial structure generated by virtual breakup, continuum propagation, and absorption in the continuum. Through the generalized optical theorem, we quantify the continuum-coupling contribution to the elastic flux loss and compare it with the elastic-breakup cross section over a broad incident-energy range. The calculation shows that the elastic-breakup fraction increases with energy, whereas the additional absorption associated with continuum components is strongest at intermediate energies.
\end{abstract}


\end{frontmatter}

\section{Introduction}
Projecting complex many-body interactions onto effective potentials in a reduced model space is a universal strategy in quantum physics, from nuclear reactions and ultracold atomic gases~\cite{Cheng10,Bloch08} to quantum optics~\cite{Theis04}. Herman Feshbach's projection-operator formalism~\cite{Feshbach58,Feshbach62} provides the rigorous foundation for this program: coupling to excluded channels generates an energy-dependent, non-local effective interaction in the retained space. While this formalism has been central to nuclear optical potentials~\cite{Mahzoon14,Mahzoon17,Idini19,Pruitt20,Whitehead21} and beyond, a key computational barrier has persisted for composite-projectile reactions: constructing the full coupled-channel Green's function in the excluded ($Q$) space requires solving coupled inhomogeneous equations with proper boundary conditions for all linearly independent solutions, which has remained a key computational barrier. As a result, practical implementations in this context have relied on either weak-coupling approximations or local equivalents that discard the non-local structure.

In the era of rare-isotope beams, exemplified by facilities such as FRIB (USA), RAON (Korea), and HIAF (China), modeling reactions with exotic, weakly bound nuclei is critical for astrophysical nucleosynthesis and medical isotope production. However, traditional phenomenological optical potentials~\cite{CH89,KD02}, parameterized from data on stable nuclei, prove inadequate when extrapolated to these unstable systems due to their limited predictive power and inability to capture the complex dynamics of weakly bound structures. As highlighted in recent reviews~\cite{Hebborn_2023}, there is a growing effort to derive optical potentials that incorporate channel-coupling effects, such as breakup and collective excitations, from underlying reaction models rather than phenomenological fits.

Previous implementations of the Feshbach effective interaction for composite-projectile reactions fall into two categories, neither of which retains the full non-local structure. (i)~Weak-coupling approaches~\cite{Potel25,Rao1973,PhysRevC.77.034609} neglect continuum-continuum couplings, making the $Q$-space Green's function diagonal; these underestimate higher-order coupling effects, whose importance in CDCC has been documented~\cite{Nunes04}. (ii)~Local-equivalent approximations such as the trivially equivalent local potential (TELP) and weighted equivalent local potential (WELP)~\cite{Canto09,rangel2025nucleusnucleuspotentialsscatteringtightly} include continuum-continuum couplings but reduce the intrinsically non-local effective interaction to a local form by dividing by the elastic wave function, introducing near-singularities at quasi-nodes and discarding the off-diagonal structure. Separately, phenomenological non-local optical potentials~\cite{PereyBuck62} assume a Gaussian form for the non-locality without connection to the underlying channel couplings. When the full wave function is expanded over channel components, the absorption functional $\langle\Psi|W|\Psi\rangle$ contains not only diagonal terms but also off-diagonal cross terms $\langle\psi_i|W_{ij}|\psi_j\rangle$ that can partially cancel the diagonal contributions~\cite{Liu2025CoherentFusion}. For weakly bound nuclei with spatially extended wave functions, a proper non-local treatment that retains this structure is essential.

For weakly bound projectiles, continuum couplings not only modify the elastic channel but also redistribute flux into various non-elastic channels~\cite{CANTO20061,CANTO20151,Jin19,cook19}. Quantifying this coupling-induced flux loss and understanding its energy dependence requires an effective interaction that retains the full channel-coupling structure, motivating the present work.

In this Letter, we present, to our knowledge, the first explicit full-coupling construction of the Feshbach dynamic polarization potential from a converged CDCC calculation, retaining the complete off-diagonal $Q$-space Green's function without weak-coupling or local-equivalent approximations.\footnote{By ``without weak-coupling or local-equivalent approximations'' we mean that all coupling terms are treated to all orders within the discretized CDCC model space, subject to standard convergence criteria.} The approach yields smooth non-local potentials whose off-diagonal structure is fully preserved. We verify the numerical implementation by confirming that the projected potential reproduces the parent CDCC elastic observables, and compare with experimental data. Through the generalized optical theorem, we quantify the coupling-induced elastic flux loss within the model space and examine its energy dependence.

\section{Theoretical framework}
We consider a two-cluster projectile $a=b+x$ incident on a target $A$. The three-body Hamiltonian is written as
\begin{equation}
H=T_A+U_{bA}+U_{xA}+h,
\label{eq:Hamiltonian}
\end{equation}
where $h=T_{bx}+V_{bx}$ is the internal Hamiltonian of the projectile, $T_A$ is the kinetic-energy operator for the projectile--target coordinate, and $U_{bA}$ and $U_{xA}$ are the fragment--target optical potentials. In the CDCC approach~\cite{jin_cdcc,AUSTERN1987125}, the projectile ground state together with the set of other projectile states, including the discretized continuum states, are denoted by $\phi_{bx}^0$ and $\phi_{bx}^i$, respectively. The unit operator is
\begin{equation}
\begin{aligned}
\mathbb{1}= &  \int R^2 d R\left|R \alpha_0\phi_{b x}^{0} \right\rangle\left\langle R \alpha_0\phi_{b x}^{0} \right| \\
& +\sum^\alpha\sum_{i=1}^{n}\int R^2\mathrm{d}R~|R\alpha\phi_{bx}^i\rangle\langle R\alpha\phi_{bx}^i|.
\end{aligned}
\end{equation}
We separate the model space into the elastic channel and the continuum subspace through
\begin{equation}
\begin{aligned}
P&=\int R^2\mathrm{d}R~|R\alpha_0\phi_{bx}^0\rangle\langle R\alpha_0\phi_{bx}^0|,\\
Q&=\sum^\alpha\sum_{i=1}^{n}\int R^2\mathrm{d}R~|R\alpha\phi_{bx}^i\rangle\langle R\alpha\phi_{bx}^i|,
\end{aligned}
\label{eq:PQ}
\end{equation}
with $P^2=P$, $Q^2=Q$, $PQ=0$, and $P+Q=1$ within the adopted CDCC model space. The channel index is denoted schematically by $\gamma=\{n,\alpha\}$, where $n$ labels the projectile internal state or bin and $\alpha$ denotes the remaining spin and orbital angular-momentum quantum numbers. The elastic channel is denoted by $\gamma_0$, for which $n=0$. For the spinless-deuteron benchmark considered here the projectile has a single bound state, so the $P$ space reduces to the single elastic channel $\alpha_0$; the general construction allows several bound projectile states to be retained in $P$.

Projecting the scattering wave function as $P|\Psi^{(+)}\rangle=|\varphi_P^{(+)}\rangle$ and $Q|\Psi^{(+)}\rangle=|\varphi_Q^{(+)}\rangle$, the Schr\"odinger equation gives, for the total energy $E$ in the center-of-mass frame,
\begin{equation}
\begin{aligned}
(E-PHP)|\varphi_P^{(+)}\rangle-PHQ|\varphi_Q^{(+)}\rangle&=0,\\
(E-QHQ)|\varphi_Q^{(+)}\rangle-QHP|\varphi_P^{(+)}\rangle&=0 .
\end{aligned}
\label{eq:projected_equations}
\end{equation}
This $E$ is the single conserved total energy, that is, the eigenvalue of the full three-body Schr\"odinger equation $(E-H)|\Psi^{(+)}\rangle=0$; the projected components $|\varphi_P^{(+)}\rangle$ and $|\varphi_Q^{(+)}\rangle$ are two parts of this single eigenstate. The total energy is not partitioned at the level of the projectors: its sharing between internal excitation and relative motion is generated by the operator $(E-QHQ)$, which assigns to a channel $\gamma$ of internal energy $\varepsilon_\gamma$ the asymptotic relative kinetic energy $E-\varepsilon_\gamma$.
Eliminating the $Q$-space component leads to an effective equation in the elastic subspace,
\begin{equation}
(E-H_{\rm eff})|\varphi_P^{(+)}\rangle=0,
\label{eq:Peq}
\end{equation}
where
\begin{equation}
\begin{aligned}
H_{\rm eff}&=PHP+U_{PQ}\mathbf{G}_{QQ}U_{QP},\\
\mathbf{G}_{QQ}(E)&=Q\frac{1}{E+i\epsilon-QHQ}Q .
\end{aligned}
\label{eq:Heff}
\end{equation}
Here $U_{PQ}=PHQ$ and $U_{QP}=QHP$. 
The first term, $PHP$, gives the bare elastic-channel interaction, while the second term is the dynamic polarization potential (DPP) induced by the eliminated continuum subspace. It represents an amplitude in which the elastic channel couples to $Q$, propagates inside the continuum space, and couples back to the elastic channel.

In an angular-momentum coupled basis, this DPP becomes a non-local kernel,
\begin{equation}
\Delta U_{\gamma_0}(R,R')=\sum_{\gamma,\gamma'\in Q}U_{\gamma_0\gamma}(R)g_{\gamma\gamma'}(R,R')U_{\gamma'\gamma_0}(R') ,
\label{eq:DPP}
\end{equation}
and the corresponding effective interaction is
\begin{equation}
U_{\rm eff}^{\gamma_0}(R,R')=U_{00}(R)\delta(R-R')+\Delta U_{\gamma_0}(R,R') .
\label{eq:Ueff}
\end{equation}
The non-locality follows directly from the $Q$-space Green's function
$g_{\gamma\gamma'}(R,R')$, which connects two different projectile--target separations through propagation in the continuum subspace. 
Thus, the DPP is not simply an additional local optical potential; it contains the spatial memory of the eliminated breakup dynamics. The kernel $U_{\rm eff}^{\gamma_0}(R,R')$ is constructed separately in each total-angular-momentum block, so it is a partial-wave-dependent non-local interaction rather than a single $J$-independent local potential.

A common simplification in earlier applications of \eqref{eq:DPP} is to neglect the coupling among different $Q$-space channels~\cite{Potel25,Rao1973,PhysRevC.77.034609}. In this weak-coupling approximation, the continuum channels communicate with the elastic channel but not with each other. The Green's function then becomes diagonal in the $Q$ space, and continuum--continuum propagation is absent. In the present work, we retain the full $Q$-space coupling. The resulting Green's function is generally non-diagonal and therefore includes both direct elastic-continuum coupling and subsequent propagation among continuum configurations. The validity of the weak-coupling approximation is governed by the strength of these continuum--continuum couplings relative to the energy denominators in $\mathbf{G}_{QQ}$, not directly by the projectile binding energy. It is reasonable only when the $Q$-space internal couplings are weak, which is not the situation for weakly bound projectiles: the spatially extended ground state and the low breakup threshold of a system such as the deuteron make the continuum--continuum couplings strong, so the full Green's function must be retained.

The coupled-channel Green's function can be obtained by several equivalent methods, such as an iterative solution or a Dyson-equation construction. Here we build it directly from the regular and irregular solutions of the coupled equations in the $Q$ space~\cite{PhysRevA.29.625,PhysRevA.30.1120,liu2026exact}; the explicit matrix construction and its uniqueness are derived in Ref.~\cite{liu2026exact}.
In compact notation,
\begin{equation}
g_{\gamma\gamma'}(R,R')=\left\{
\begin{aligned}
\frac{2\mu}{\hbar^2}\sum_m^{Q}\frac{u_\gamma^m(R)h_{\gamma'}^m(R')}{\mathcal{W}^m(R')} ,&\qquad R<R' , \\
\frac{2\mu}{\hbar^2}\sum_m^{Q}\frac{h_{\gamma}^m(R)u_{\gamma'}^m(R')}{\mathcal{W}^m(R')} ,&\qquad R>R' .
\end{aligned}
\right.
\label{eq.Green}
\end{equation}
Here $u^m$ and $h^m$ are the linearly independent regular and irregular (outgoing) solutions of the coupled $Q$-space equation $(E-QHQ)\chi=0$ at the same total energy $E$, the index $m$ labelling the independent solutions of this coupled system, and $\mathcal{W}^m$ denotes the corresponding Wronskian. The derivative discontinuity of \eqref{eq.Green} at $R=R'$ produces the delta function required by the coupled radial equations, while the regular and outgoing solutions impose the proper boundary conditions. \eqref{eq.Green} is therefore the practical step that converts the formal Feshbach expression into a calculable non-local DPP. If the off-diagonal elements of the $Q$-space Hamiltonian are removed, \eqref{eq.Green} reduces to the weak-coupling limit. Once the kernel $\Delta U_{\gamma_0}(R,R')$ is built, the non-local effective equation~\eqref{eq:Peq} is solved with the $R$-matrix method on a Lagrange mesh~\cite{Descouvemont_2010}, the integral kernel is evaluated by Gauss--Lagrange quadrature and treated on the same footing as the local part, so that the elastic $S$ matrix follows directly from a single linear solve, without iteration~\cite{liu2026exact}.

The reaction flux associated with the continuum-induced DPP is obtained from the generalized optical theorem~\cite{Cotanch10,Liu2025CoherentFusion,Thompson_Nunes_2009},
\begin{equation}
\sigma_{\rm DPP}=\frac{2}{\hbar v_0}\frac{4\pi}{k_0^2}\sum_{\gamma_0}\Im\left\langle\varphi_P^{\gamma_0}\left|\Delta U_{\gamma_0}\right|\varphi_P^{\gamma_0}\right\rangle ,
\label{eq:sigmaDPP}
\end{equation}
where $v_0$ and $k_0$ are the incident velocity and wave number in the elastic channel. For real fragment--target interactions, this quantity reduces to the elastic-breakup contribution generated by the continuum channels. For complex fragment--target optical potentials, however, the continuum components can also be absorbed. Therefore, $\sigma_{\rm DPP}$ contains not only elastic breakup but also absorption associated with breakup configurations, including nonelastic breakup and breakup-fusion pathways~\cite{Jin19,PhysRevC.108.034612}. 

This point is important for weakly bound projectiles, because breakup and absorption are strongly coupled and do not occur as two clearly separate steps. In channel notation, the absorptive part contains matrix elements of the imaginary interaction, including cross terms such as $\langle\psi_i|W_{ij}|\psi_j\rangle$. These off-diagonal terms are part of the flux-conserving CDCC absorption functional and describe interference among continuum pathways~\cite{Liu2025CoherentFusion}. Thus, the same full Green's function that generates the non-local DPP also carries the continuum-interference effects that enter the evaluation of $\sigma_{\rm DPP}$.

\section{Results}
We apply the method to deuteron scattering from $^{58}$Ni at incident energies from 20 to 80~MeV. The deuteron is described by a Gaussian central $p$-$n$ interaction reproducing the 2.225~MeV binding energy~\cite{AUSTERN1987125}; tensor components and intrinsic spins are not included in the present calculation. This simplified structure makes the deuteron an ideal first benchmark because there are no bound excited states and the breakup continuum is well defined. The nucleon-target optical potentials are taken from the KD02 global parameterization~\cite{KD02}, each evaluated at half the deuteron incident energy and treated as fixed input operators in the Feshbach reduction. The CDCC model space includes $p$-$n$ continuum states up to internal orbital angular momentum $\ell\leq4$, with sufficiently extended energy bins, and the projectile-target total angular momentum is truncated at $J_{\rm max}=120$.

The convergence of the extracted effective interaction is tied to the convergence of the parent CDCC calculation. We checked the stability against $J_{\rm max}$, continuum-bin density, maximum continuum energy, and $p$-$n$ partial-wave truncation. Increasing the model-space parameters beyond the adopted values changes the relevant cross sections by less than about 1\%. The breakup cross section is more sensitive to the $p$-$n$ partial-wave cutoff than the elastic angular distribution, but the elastic observables entering the projected two-body potential are already stable in the adopted space. This distinction is useful: the non-local potential is fixed by the elastic-channel projection, whereas the breakup observables test how much continuum strength is explicitly resolved. In the figures below the same CDCC model space is used for both the many-channel calculation and the projected effective interaction, so any difference between them would indicate an error in the construction of $g_{\gamma\gamma'}$ or in the numerical evaluation of the DPP kernel.

\begin{figure}[h]
    \centering
    \includegraphics[width=1.0\linewidth]{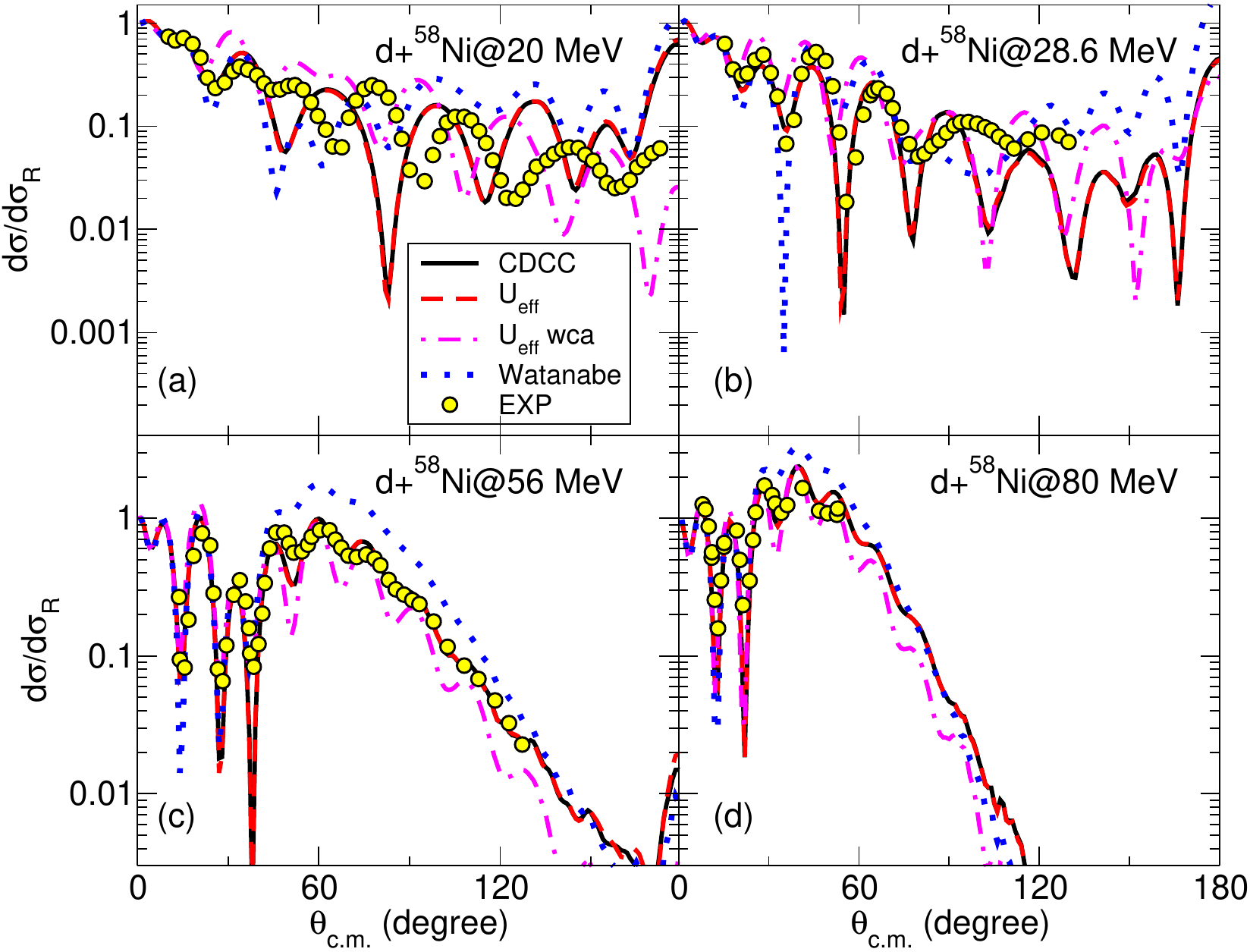}
    \caption{Elastic scattering angular distributions for $d+^{58}$Ni. Solid lines: CDCC; dashed lines: full-coupling $U_{\rm eff}$; dot-dashed lines: weak-coupling result; dotted lines: bare $U_{00}$ folding baseline (Watanabe). Experimental data are shown for $E_d=20$~MeV~\cite{20ERMER199171}, 28.6~MeV~\cite{28_6PERRIN1977221}, 56~MeV~\cite{56MATSUOKA1986413}, and 80~MeV~\cite{DUHAMEL1971485}.}
    \label{fig:elastic}
\end{figure}

Figure~\ref{fig:elastic} compares the elastic angular distributions obtained from the original CDCC calculation, the full-coupling effective potential, the weak-coupling effective potential, and the Watanabe folding potential~\cite{Watanabe_pot}. The latter is the single-folding of the same KD02 nucleon optical potentials~\cite{KD02} over the deuteron ground state, so it coincides with the bare term $U_{00}(R)$ in \eqref{eq:Ueff}, that is, the elastic-channel interaction before the continuum-induced DPP is added. It is therefore built from the same interactions as the CDCC calculation and serves as a no-DPP baseline rather than an independent phenomenological potential. The full-coupling $U_{\rm eff}$ closely follows the CDCC result over the whole angular range and for all the energies shown. This agreement shows that the effective two-body nonlocal interaction obtained from the Green's-function construction retains the elastic-scattering information of the original CDCC calculation. The comparison with the data shows that the underlying CDCC calculation based on the KD02 nucleon optical potential gives a reasonable description of this benchmark reaction, while the agreement between CDCC and $U_{\rm eff}$ tests the consistency of the Feshbach projection itself. By contrast, the weak-coupling result and the bare folding (Watanabe) baseline both deviate visibly from CDCC, consistent with the long-known inadequacy of Watanabe folding for $d+^{58}$Ni~\cite{AUSTERN1987125}; here this deviation serves as an orientation baseline that sets the scale of the continuum-induced DPP.

\begin{figure}[h]
    \centering
    \includegraphics[width=1.0\linewidth]{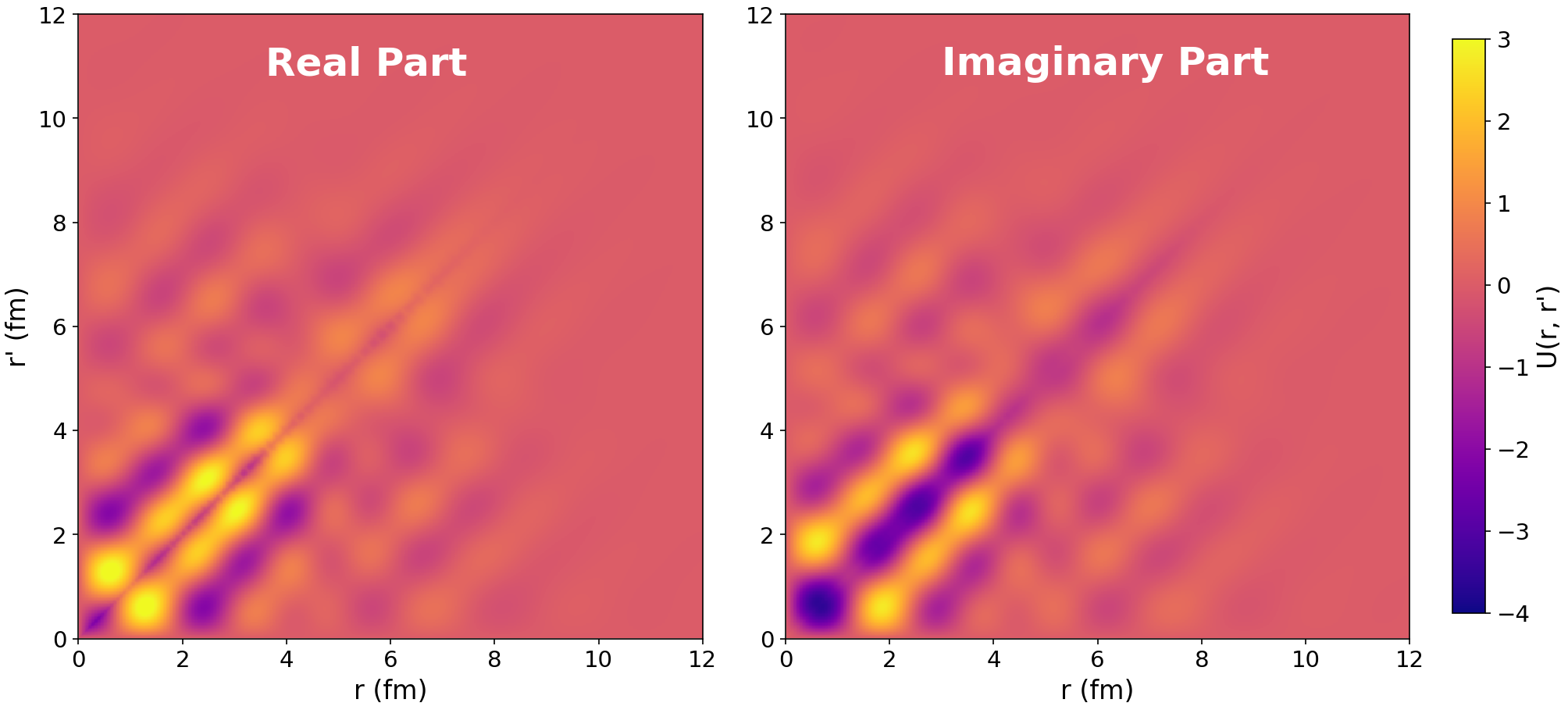}
    \caption{Real and imaginary parts of the non-local DPP $\Delta U(R,R')$ for $d+^{58}$Ni at 56~MeV and $J=0$. The strength is concentrated near the diagonal but extends over a finite off-diagonal range.}
    \label{fig:nonlocal}
\end{figure}

Figure~\ref{fig:nonlocal} reveals the physical content that local approximations cannot properly represent. The DPP exhibits rich off-diagonal structure, the spatial counterpart of the continuum-continuum couplings retained in $g_{\gamma\gamma'}$. Within the Feshbach-CDCC framework, the coupling-induced flux redistribution is encoded in the spatial structure of the non-local potential and constrained by unitarity. While TELP may reproduce correct cross sections numerically, it is mathematically ill-defined at the quasi-nodes of the wave function (where the required division becomes singular) and therefore cannot provide a rigorous representation of this structure.

\begin{figure}[h]
    \centering
    \includegraphics[width=1.0\linewidth]{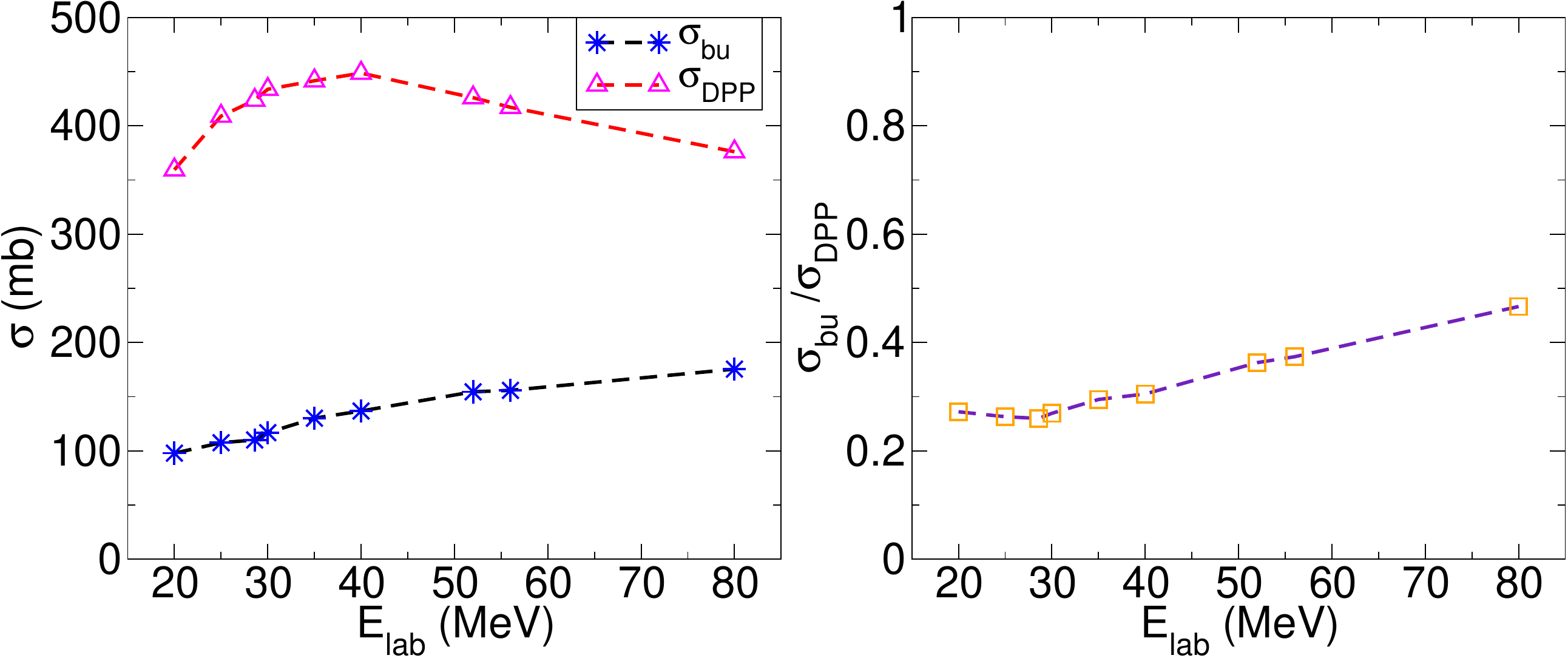}
    \caption{Left: elastic-breakup cross section $\sigma_{bu}$ and DPP-induced reaction contribution $\sigma_{\rm DPP}$ as functions of incident energy. Right: ratio $\sigma_{bu}/\sigma_{\rm DPP}$.}
    \label{fig:mechanism}
\end{figure}

To explore the energy dependence of coupling effects, Fig.~\ref{fig:mechanism} compares the elastic breakup cross section $\sigma_{bu}$ with $\sigma_{\rm DPP}$ defined in \eqref{eq:sigmaDPP} over a range of incident energies. Here $\sigma_{\rm DPP}$ quantifies the total elastic flux loss induced by continuum couplings within the CDCC model space, while $\sigma_{bu}$ is calculated from the CDCC $S$-matrix elements connecting the ground state to asymptotic continuum channels. The difference $\sigma_{\rm DPP}-\sigma_{bu}$ therefore represents the additional absorption arising from the imaginary parts of the nucleon--target optical potentials acting on the continuum components of the wave function~\cite{Jin19,PhysRevC.108.034612}. By construction this difference is an absorption cross section and is non-negative, so $\sigma_{\rm DPP}\ge\sigma_{bu}$ at all energies considered. The elastic-breakup contribution increases smoothly with incident energy, whereas the DPP-induced flux loss is non-monotonic: it grows at low energies, reaches its largest values in the intermediate-energy region, and then decreases slightly at higher energies. As a result, the ratio $\sigma_{bu}/\sigma_{\rm DPP}$ increases with energy, indicating that elastic breakup becomes a larger fraction of the continuum-coupling effect at higher energies, while continuum absorption is relatively more important near the intermediate-energy region. The result also clarifies the interpretation of the DPP cross section: it should not be identified with elastic breakup alone once the fragment-target interactions are complex. In that case, elastic breakup is the part of the continuum flux that survives asymptotically as breakup fragments, while $\sigma_{\rm DPP}$ counts the total loss from the elastic channel induced by coupling to the continuum, including the part subsequently absorbed by fragment-target optical potentials. This distinction is especially important for weakly bound projectiles, because breakup and absorption are strongly coupled and do not occur as two clearly separate steps.

\section{Conclusion}
We have constructed the Feshbach dynamic polarization potential from a full-coupling CDCC calculation by retaining the complete $Q$-space Green's function. Within the adopted discretized model space, the resulting non-local effective interaction reproduces the CDCC elastic observables and therefore provides a phase-equivalent two-body representation of the coupled three-body dynamics.

Application to $d+^{58}$Ni shows that continuum-continuum propagation affects the extracted effective interaction and that the DPP has a smooth, finite-range non-local structure. The generalized optical theorem further connects this non-local interaction to reaction flux loss. For real fragment-target interactions, this flux corresponds to elastic breakup; for complex fragment-target optical potentials, it also contains absorption from continuum components. The observed energy dependence shows a rising elastic-breakup fraction and a non-monotonic continuum-absorption contribution.

The construction is demonstrated here for deuteron breakup within CDCC. Extension to other coupled-channels systems (collective excitations, transfer couplings) is possible in principle within the same Feshbach projection framework, though each case will require its own convergence study.

The extracted phase-equivalent two-body potential encodes the full coupled-channels dynamics while preserving quantum coherence effects. A natural next step is to employ this potential in distorted-wave Born approximation (DWBA) calculations for transfer reactions, which would incorporate continuum-coupling effects at reduced computational cost; this application remains to be demonstrated.

By eliminating the mathematical singularities inherent in local equivalent potentials, the approach also enables quantitative study of how continuum couplings modify the effective nuclear radius and diffuseness. Notably, the explicitly constructed non-local potential (Fig.~\ref{fig:nonlocal}) reveals that the non-local structure generated by channel couplings is not constrained to the Gaussian form commonly assumed since the work of Perey and Buck~\cite{PereyBuck62}. Future investigations include the energy dependence and dispersion relation of the effective potential and applications to halo nuclei.

\section*{Acknowledgements}
This work was supported by the National Natural Science Foundation of China (Grant No.~12535009 and No.~12475132), by the National Key R\&D Program of China (Contracts No. 2023YFA1606503), and by the Fundamental Research Funds for the Central Universities.

\bibliography{cdcc}

\end{document}